\documentclass[
    ,final            
  ]
  {aipproc}

\layoutstyle{8x11double}

\newcommand{\be}{\begin{equation}}
\newcommand{\ee}{\end{equation}}
\newcommand{\bea}{\begin{eqnarray}}
\newcommand{\eea}{\end{eqnarray}}
\newcommand{\bt}{\begin{tabular}}
\newcommand{\et}{\end{tabular}}
\newcommand{\etal}{{\it et al.}}

\begin{document}

\title{Neutrinos and Cosmology: an update}


\author{Ofelia Pisanti}{
  address={Dipartimento di Scienze Fisiche, Universit\`{a} di
  Napoli {\it Federico II}, and INFN, Sezione di Napoli,
  \\Complesso Universitario di Monte Sant'Angelo, Via Cintia,
  I-80126 Napoli, Italy} }

\author{Pasquale D. Serpico}{
  address={Max Planck Institut f\"{u}r Physik,
  Werner-Heisenberg-Institut,
  \\F\"{o}hringer Ring 6, 80805, M\"{u}nchen, Germany}
}

\begin{abstract}
We review the current cosmological status of neutrinos, with
particular emphasis on their effects on Big Bang Nucleosynthesis,
Large Scale Structure of the universe and Cosmic Microwave
Background Radiation measurements.
\end{abstract}

\maketitle


\section{Introduction}

The Standard Cosmological Model predicts the existence of a
neutrino background (C$\nu$B) filling the universe with densities
of the order $n_\nu\approx 110\:{\rm cm}^{-3}$ per flavor.
Neutrino properties are rather difficult to be probed
experimentally, due to the weakness of neutrino interactions
which, especially at low energies, makes hopeless at present any
perspective of direct detection of the C$\nu$B. Nevertheless,
neutrinos are one of the most abundant relics of the primordial
universe and played a key role in different stages of its
evolution. Several cosmological observables are then sensitive to
neutrinos, and can be used to put bounds on their properties.

Given their extremely low interaction rate, the natural
out-of-equilibrium driving force of the expansion of the Universe
pushed neutrinos to decouple from the thermal bath very early,
when the temperature was ${\cal O}$(1~MeV). This temperature is
close to the electron mass $m_e$, setting the scale of the
electron/positron annihilation, and both are close to the ${\cal
O}$(0.1~MeV) scale of the synthesis of the light nuclei via
thermonuclear fusion. So, Big Bang Nucleosynthesis (BBN) is a
privileged laboratory for the C$\nu$B studies. In particular, it
is sensitive to the $\nu$ (weak) interactions as well as to the
shape of the $\nu_e-\bar{\nu}_e$ phase space distributions
entering the $n\leftrightarrow p$ inter-conversion rates. Apart
from the energy density due to the extra ({\it i.e.} non
electromagnetic) relativistic degrees of freedom, the BBN tests
the \emph{dynamical} properties of the neutrinos in a thermalized
(almost) CP-symmetric medium.

Other cosmological probes are Cosmic Microwave Background (CMB)
anisotropies or the Large Scale Structure (LSS) of the universe,
which are, however, sensitive only to the C$\nu$B
\emph{gravitational} interaction. The role of neutrinos as the
dark matter (DM) particles has been widely discussed since the
early 1970s. For values of neutrino masses much larger than the
present cosmic temperature one finds a contribution in terms of
the critical density $\Omega_\nu \simeq 0.0108\,h^{-2}\sum
m_\nu/$eV, $h$ being the Hubble parameter in units of 100 Km
s$^{-1}$ Mpc$^{-1}$. Nowadays, we know that neutrinos cannot
constitute all the DM ($\Omega_{\rm DM}\approx\Omega_{m}\sim 0.3$
\cite{Spergel:2003cb}), and the main question is how large the
contribution of neutrinos can be, deducing $\Omega_\nu$ from their
contribution to cosmological perturbations. In fact, neutrino
background erases the density contrasts on wavelengths smaller
than a mass-dependent free-streaming scale. Neutrinos of sub-eV
mass behave almost like a relativistic species for CMB
considerations and therefore the power spectrum suppression can be
seen only in LSS data. Even if neutrino mass influences only
slightly the spectrum of CMB anisotropies, it is crucial to
combine CMB and LSS observations, because the former give
independent constraints on the cosmological parameters, and
partially remove the parameter degeneracy that would arise in an
analysis of the LSS only.

\section{Neutrinos and cosmology}

At temperatures above ${\cal O}$(1~MeV), neutrinos are in thermal
equilibrium with the thermal bath and their distribution is a
perfect Fermi-Dirac one,
\be
f_{\nu_\alpha} (y) = \frac{1}{e^{y-\xi_\alpha}+1},
\label{eq:fnu}
\ee
where $y\equiv p/T_\nu$ and $\xi_\alpha \equiv \mu_\alpha/T_\nu$
(here $\mu_\alpha$ is the chemical potential of the flavor
$\alpha$, which is neglected in the standard scenario). As the
temperature goes down, the universe expansion prevents weak
interactions from maintaining neutrinos in equilibrium and they
decouple. As a first approximation, the neutrino decoupling can be
described as an instantaneous process taking place around 2-4 MeV,
without any overlap in time with $e^{+}-e^{-}$ annihilation. All
flavors would then keep Fermi-Dirac distributions (both neutrino
momenta and temperature redshift identically with the universe
expansion), but the neutrino temperature $T_\nu$ will not benefit
of the entropy release from $e^{+}-e^{-}$ annihilations. The
asymptotic ratio $T/T_\nu$ for $T\ll m_e$ can be evaluated in an
analytic way, and turns out to be $(11/4)^{1/3}\approx 1.401$.

More accurate calculations by solving the kinetic equations have
been performed, and they show a partial entropy transfer to the
neutrino plasma. As a consequence, the neutrino distributions get
distorted, since this transfer is more efficient for larger
neutrino momenta. In \cite{Mangano:2001iu,Esposito:2000hi} it was
shown that with a very good approximation the distortion in the
$\alpha$-th flavor can be described as
\be
f_{\nu_\alpha} \left( x,y\right) \simeq \frac{1}{e^{y }+1} \left(
1 + \sum_{i=0}^3\, c_{i}^\alpha(x)\, y^i\right),
\ee
with $x\equiv m_e/T_\nu$. The electron neutrinos get a larger
entropy transfer than the $\mu$ and $\tau$, since they also
interact via charged currents with the $e^{\pm}$ plasma. The
effective ratio $T/T_\nu\approx 1.3984$ is slightly lower than the
instantaneous decoupling estimate.

The incomplete decoupling of neutrinos also induces a modification
in the contribution of neutrinos to the energy density. By fully
consistently including order $\alpha$ QED corrections to the
photon and $e^{\pm}$ equation of state, in \cite{Mangano:2001iu}
the energy density in the neutrino fluid is found to be enhanced
by 0.935\% (for $\nu_e$) and 0.390\% (for $\nu_\mu$ and
$\nu_\tau$). A refined treatment, also including the effects of
three-flavor neutrino oscillations, has been recently provided in
\cite{Mangano:2005cc}.

The contribution of neutrinos to the total relativistic energy
density of the universe is usually parameterized via the
``effective number'' of neutrinos, ${\rm N}_{\rm eff}$,
\be
\rho_\nu = {\rm N}_{\rm eff}~ \frac{7}{8}\, \left( \frac{T_\nu}{T}
\right)^4\, \rho_\gamma.
\label{eq:energydensity}
\ee
${\rm N}_{\rm eff}$ measures neutrino energy density in ``units''
of the energy density of massless neutrinos with zero chemical
potential, but it can in principle receive a contribution from
other (relativistic) relics. For three massless neutrinos with
zero chemical potential and in the limit of instantaneous
decoupling, ${\rm N}_{\rm eff}=3$. The inclusion of entropy
transfer between neutrinos and the thermal bath modifies this
number to about 3.04 at the CMB epoch.

Shortly after neutrino decoupling the temperature reaches the
value of the neutron-proton mass difference, and weak interactions
are no longer fast enough to maintain equilibrium among nucleons:
a substantial final neutron fraction survives, however, down to
the phase of nucleosynthesis where all neutrons become practically
bound in $^{4}$He nuclei. The predicted value of the $^4$He mass
abundance, $Y_p$, is poorly sensitive to the nuclear network
details and has only a weak, logarithmic dependence on the baryon
fraction of the universe, $\omega_b= \Omega_b h^2$, being fixed
essentially by the ratio of neutron to proton number density at
the onset of nucleosynthesis. This in turn crucially depends on
the weak rates and on the (standard or exotic) neutrino
properties.

\begin{figure}
\includegraphics[height=.3\textheight]{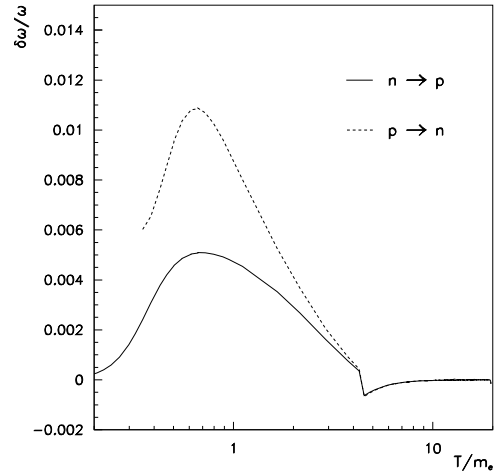}
\caption{The relative correction to the $n \rightarrow p$ (solid
line) and $p \rightarrow n$ (dashed line) total rates, due to the
neutrino distortion (see Ref. \cite{Serpico:2004gx} for details).}
\label{fig:newcorrnu}
\end{figure}

In several papers (see \cite{Lopez:1998vk,Esposito:1999sz} and
references therein) the value of $Y_p$ has been computed by
improving the evaluation of the weak rates including
electromagnetic radiative corrections, finite mass corrections and
weak magnetism effects, as well as the plasma and thermal
radiative effects. In particular in \cite{Serpico:2004gx} it has
been also considered the effect of the neutrino spectra
distortions and of the process $\gamma+p \leftrightarrow \nu_e +
e^{+}+ n$, which is kinematically forbidden in vacuum, but allowed
in the thermal bath. The latter is shown to give a negligible
contribution, while neutrino distortions have a significant
influence on the rates for different reasons: a) the larger mean
energy of $\nu_e$ induces a $\delta T_{\nu_e}$ and (indirectly,
through a decrease in $\rho_{e.m.}$) a $\delta T$; b) the ratio
$T/T_{\nu}$, which enters the weak rates, is changed. Moreover,
the time-temperature relationship is changed and so the time at
which BBN starts.

The total effect on the rates are shown in Figure
\ref{fig:newcorrnu}. Even though one would expect effects up to
${\cal O}$(1\%), the spectral distortion and the changes in the
energy density and $T_\nu(T)$ conspire to almost cancel each
other, so that $Y_p$ is changed by a sub-leading ${\cal
O}$(0.1\%). This effect is of the same order of the predicted
uncertainty coming from the error on the measured neutron
lifetime, $\tau_n=885.7\pm 0.8$ s \cite{Eidelman:2004wy}, and has
to be included in quoting the theoretical prediction,
$Y_p=0.2481\pm 0.0004\:(1\,\sigma$, for $\omega_b=0.023\pm
0.001$).

Apart from the uncertainty on $\omega_b$, the $^2$H, $^3$He and
$^7$Li abundance predictions are mainly affected by the nuclear
reaction uncertainties. An updated and critical review of the
nuclear network and a new protocol to perform the nuclear data
regression has been widely discussed in \cite{Serpico:2004gx}, to
which we address for details.

A different scenario, the Degenerate BBN (DBNN) one, has received
a new attention in the last few years, especially when the first
data on CMB seemed to indicate a tension between the determination
of $\omega_b$ from CMB and standard BBN \cite{Esposito:2000hh}.
Such a tension could, in fact, be relaxed assuming that the number
of $\nu_\alpha$ and  ${\bar\nu}_\alpha$ be different, i.e.
$\mu_\alpha\neq 0$ in Eq.\eqref{eq:fnu}. In this degenerate case,
changes are expected: a) in the weak rates (a reduction in $Y_p$),
since a positive $\xi_e$ enhances $n \rightarrow p$ processes with
respect to the inverse ones, and modifies the initial condition on
the $n/p$ ratio at $T\gg$ 1 MeV; b) in the expansion rate (an
increase in $Y_p$), since non zero $\xi_\alpha$'s contribute to
total ${\rm N}_{\rm eff}$ as
\be
{\rm N}_{\rm eff}^{\rm DBBN} \simeq {\rm N}_{\rm eff} +
\sum_\alpha \left[ \frac{30}{7}
\left(\frac{\xi_\alpha}{\pi}\right)^2 + \frac{15}{7}
\left(\frac{\xi_\alpha}{\pi}\right)^4 \right].
\label{eq:rhonudeg}
\ee

When several neutrino species are degenerate, both the effects
might combine and particular values of $\xi_\alpha$'s exist for
which the predictions of DBBN are still in good agreement with the
observational data on the abundances of primordial elements.
Notice that BBN is more sensitive to neutrino degeneracy than CMB
or LSS, due to the further effect played by the
$\nu_e-\bar{\nu}_e$ distributions in the weak rates, while the
latter are only sensitive to the extra energy density present in
the $\xi\neq 0$ case.

Earlier claims of discrepancies between BBN and CMB have been
largely overcome by new data, and it was recently realized
\cite{Dolgov:2002ab} that the flavor oscillations in the
primordial plasma induced by (presently determined) mass
differences and mixing angles from atmospheric and solar neutrinos
almost equalize the three asymmetry parameters $\xi_\alpha$.
Still, exotic models, where both a common relatively large $\xi$
and a ${\rm N}_{\rm eff}\neq 3.04$ exist, have been considered
(``hidden relativistic degrees of freedom''
\cite{Barger:2003rt,Cuoco:2003cu}), and were shown to be
compatible with the data. However, if one sticks to the scenario
with ${\rm N}_{\rm eff}$ fixed by standard Physics, introduces no
sterile species, and assumes the CMB prior on $\omega_b$, BBN
turns into a poweful "leptometer", constraining the common $\xi$
to an unprecedent accuracy even under conservative assumptions for
$Y_p$ \cite{Serpico:2005bc}. This provides an indirect consistency
check for the sphaleron mechanism at electroweak phase transition,
predicting baryon and lepton asymmetries of the same order.

Let us come to neutrino role in structure formation. In general,
neutrinos tend to stream freely across gravitational potential
wells, and to erase density perturbations. Free-streaming is
efficient on a characteristic scale $\lambda_{\rm J}$ called the
Jeans length, corresponding roughly to the distance on which
neutrinos can travel in a Hubble time. For ultra-relativistic
neutrinos, $\lambda_{\rm J}$ is by definition equal to the Hubble
radius $c/H$, but for non-relativistic ones it is lower than
$c/H$. Neutrinos with masses smaller than approximately 0.3 eV are
still relativistic at the time of last scattering, and their
direct effect on the CMB perturbations is identical to that of
massless neutrinos. In the intermediate mass range from $10^{-3}$
eV to 0.3 eV, the transition to the non-relativistic regime takes
place during structure formation, and the matter power spectrum
will be directly affected in a mass-dependent way. Wavelengths
$\lambda$ smaller than the current value of the neutrino
$\lambda_{\rm J}$ are suppressed by free-streaming. The largest
wavelengths, which remain always larger than the neutrino
$\lambda_{\rm J}$, are not affected. Finally, there is a range of
intermediate $\lambda$ which become smaller than the neutrino
$\lambda_{\rm J}$ for some time, and then encompass it again:
these scales smoothly interpolate between the two regimes. The net
signature in the matter power spectrum $P(\lambda,\tau)$ is a
damping of all wavelengths smaller than the Hubble scale at the
time $\tau_0$ of the transition of neutrinos to a non-relativistic
regime.

Then, for $\lambda\ll\lambda_{\rm J}$, if $\Omega_\nu \ll
\Omega_m$ the suppression is given roughly by the factor
\be
\frac{\Delta P}{P} \simeq -8~ \frac{\Omega_\nu}{\Omega_m},
\ee
that is by the ratio between neutrino and matter energy densities.

Notice that one can somehow play with both ${\rm N}_{\rm eff}$ and
$\sum m_\nu$ and find models which give excellent fits of the
data. In fact, models with massive neutrinos have suppressed power
at small scale. Adding relativistic energy further suppresses
power at scales smaller than the horizon at matter-radiation
equality. For the same matter density such a model would therefore
be even more incompatible with data. However, if the matter
density is increased together with $m_\nu$ and ${\rm N}_{\rm
eff}$, data can be described very nicely (see, for example Figure
3 in \cite{Hannestad:2004nb}).

\section{Comparison with data}

Still few years ago, the BBN theory together with the observations
of the abundances of primordial nuclides were used to determine
the baryon fraction of the universe, $\omega_b$. Nowadays
$\omega_b\approx 0.023$ is fixed to better than 5\% accuracy by
detailed CMB anisotropies analysis \cite{Spergel:2003cb}, thus
leaving the BBN as an over-constrained and very predictive theory.
Once $\omega_b=0.023 \pm 0.001$ is plugged into the BBN theory,
the prediction for the deuterium, which is the nuclide most
sensitive to $\omega_b$, nicely fits the range of the observed
values in high redshift, damped Ly-$\alpha$ QSO systems
\cite{Kirkman:2003uv}, thus offering a remarkable example of
internal consistency of the current cosmological scenario.
Moreover, the predictions of other light nuclei which at least
qualitatively agree with the observed values are likely to put
constraints on the Galactic chemical evolution ($^3$He) or on the
temperature scale calibration or depletion mechanisms in PopII
halo stars ($^7$Li).

The determination of $Y_p$ is usually performed by extrapolating
to zero metallicity the measurements done in dwarf irregular and
blue compact galaxies. The typical statistical errors are of the
order of 0.002 ({\it i.e.}, at the 1\% level), but the systematics
are such that in the recent reanalysis \cite{Olive:2004kq} the
authors argue for the conservative range $0.232\leq Y_p\leq
0.258$, {\it i.e.} a 1 $\sigma$ error of ${\cal O}$(5\%).
\begin{table}
\begin{tabular}{ccl}
\hline \tablehead{1}{c}{b}{Ref.}
  & \tablehead{1}{c}{b}{Bound on ${\rm N}_{\rm eff}$}
  & \tablehead{1}{l}{b}{Data used} \\
\hline

\cite{Serpico:2004gx}   & $1.8\leq {\rm N}_{\rm eff}\leq 3.7$           & CMB,BBN\\
\cite{Cuoco:2003cu}     & $1.3\leq {\rm N}_{\rm eff}\leq 6.1$           & CMB, BBN(D) \\
\cite{Cuoco:2003cu}     & $1.6\leq {\rm N}_{\rm eff}\leq 3.6$           & BBN(D+$Y_p$) \\
\cite{Crotty:2003th}    & $1.4\leq {\rm N}_{\rm eff}\leq 6.8$           & CMB, LSS, HST \\
\cite{Hannestad:2003xv} & $1.9(2.3)\leq {\rm N}_{\rm eff}\leq 7.0(3.0)$ & CMB, LSS, (+BBN) \\
\cite{Barger:2003zg}    & $1.7\leq {\rm N}_{\rm eff}\leq 3.0$           & CMB, BBN \\
\cite{Cyburt:2004yc}    & ${\rm N}_{\rm eff}\leq 4.6$                   & CMB, BBN \\
\cite{Pierpaoli:2003kw} & $1.90\leq {\rm N}_{\rm eff}\leq 6.62$         & CMB, LSS, HST \\
\hline
\end{tabular}
\caption{Bounds on ${\rm N}_{\rm eff}$ (2 $\sigma$) from different analyses}
\label{tab:neff}
\end{table}

In Table \ref{tab:neff} the present bounds on the effective number
of neutrinos from various analyses are presented, together with
the type of data used. The most stringent bounds come from BBN
alone (Deuterium+Helium), while CMB and BBN-Deuterium are less
effective in constraining ${\rm N}_{\rm eff}$. Some differences
are due to (slightly) different databases or assumptions.

\begin{table}[b]
\begin{tabular}{ccl}
\hline \tablehead{1}{c}{b}{Ref.}
  & \tablehead{1}{c}{b}{Bound on $\xi_e$}
  & \tablehead{1}{l}{b}{Data used} \\
\hline
\cite{Barger:2003rt}  & $-0.10\leq \xi_e\leq 0.25$ & CMB, BBN \\
\cite{Cuoco:2003cu}   & $-0.13\leq \xi_e\leq 0.31$ & BBN(D+$Y_p$)+prior on $\omega_b$ \\
\cite{Serpico:2005bc} & $-0.05\leq \xi_e\leq 0.07$ & BBN($Y_p$)+priors on $\omega_b,{\rm N}_{\rm eff}$ \\
\hline
\end{tabular}
\caption{Bounds on $\xi_e$ (2 $\sigma$) from different analyses}
\label{tab:xie}
\end{table}

Table \ref{tab:xie} shows the bounds on $\xi_e$. The interval from
Ref. \cite{Cuoco:2003cu} is broader than what in Ref.
\cite{Barger:2003rt}, since in the first case only a prior from
CMB instead of all data is used in the analysis. The third line
shows the bound obtained in \cite{Serpico:2005bc} assuming only
standard physics, while the previous two bounds assume no prior on
${\rm N}_{\rm eff}$.

\begin{table}
\begin{tabular}{cll}
\hline \tablehead{1}{c}{b}{Ref.}
  & \tablehead{1}{l}{b}{Bound on $\sum m_\nu$}
  & \tablehead{1}{l}{b}{Data used} \\
\hline
\cite{Spergel:2003cb}   & $\leq 0.69$             & CMB, LSS, Ly$\alpha$ \\
\cite{Hannestad:2003xv} & $\leq 1.01$             & CMB, LSS, HST, SNIa\\
\cite{Hannestad:2004bu} & $\leq 0.65$             & CMB, LSS, HST, Ly$\alpha$ \\
\cite{Allen:2003pt}     & $=0.56_{-0.26}^{+0.30}$ & CMB, LSS, f$_{gas}$, XLF \\
\cite{Tegmark:2003ud}   & $\leq 1.7$              & CMB, LSS \\
\cite{Barger:2003vs}    & $\leq 0.75$             & CMB, LSS, HST \\
\cite{Crotty:2004gm}    & $\leq 1.0 (0.6)$        & CMB, LSS (+HST, SNIa)\\
\cite{Fogli:2004as}     & $\leq 0.47 $            & CMB, LSS, HST, SNIa, Ly$\alpha$\\
\hline
\end{tabular}
\caption{Bounds on $\sum m_\nu$ (2 $\sigma$) from different analyses}
\label{tab:summ}
\end{table}

Finally, Table \ref{tab:summ} shows, with the same notations, the
upper bound on the neutrino mass. As can be gauged from this
table, a fairly robust bound on the sum of neutrino masses is at
present somewhere around 1 eV, depending on the specific priors
and data sets used.\\{}\\
In conclusion, the complementarity among different fields of
cosmology (BBN, CMB, LSS) can be used to test the role of
neutrinos from very early epochs (redshift $z\simeq 10^{10}$) down
to relatively recent history ($z\simeq $ a few). New nuclear rate
measurements and a better understanding of possible systematics
affecting primordial abundance determination (BBN) and more data
together with an increased precision (CMB and LSS) will give us
the opportunity to constrain standard neutrino properties as well
as to test new physics in the neutrino sector, gaining at the same
time a deeper insight on the physics of the early universe.


\end{document}